\documentclass[conference,peer-reviewed]{aesconf}

\graphicspath{{./}{figures/}}

\usepackage[utf8]{inputenc}

\usepackage{microtype}

\usepackage[numbers,square]{natbib}

\usepackage{amsmath,amssymb,amsfonts,multirow,makecell}
\usepackage{algorithmic}
\usepackage{graphicx}
\usepackage{textcomp}
\usepackage{xcolor}

\usepackage{booktabs}
\usepackage{color}
\usepackage{url}
\usepackage[utf8]{inputenc}

\title{AudioGAN: A Compact and Efficient Framework for Real-Time High-Fidelity Text-to-Audio Generation}

\author[1]{HaeChun Chung}

\affil[1]{KT Corporation, Seoul, Republic of Korea}

\correspondence{HaeChun Chung}{hc.chung@kt.com}

\lastnames{HaeChun Chung}

\shorttitle{AudioGAN}

\begin{document}

\twocolumn[
\maketitle 

\begin{onecolabstract}
Text-to-audio (TTA) generation can significantly benefit the media industry by reducing production costs and enhancing work efficiency. However, most current TTA models—primarily diffusion-based—suffer from slow inference speeds and high computational costs. In this paper, we introduce AudioGAN, the first successful Generative Adversarial Networks (GANs)-based TTA framework that generates audio in a single pass, thereby reducing model complexity and inference time. To overcome the inherent difficulties in training GANs, we integrate multiple ,contrastive losses and propose innovative components—Single-Double-Triple (SDT) Attention and Time-Frequency Cross-Attention (TF-CA). Extensive experiments on the AudioCaps dataset demonstrate that AudioGAN achieves state-of-the-art performance while using 90\% fewer parameters and running 20 times faster, synthesizing audio in under one second. These results establish AudioGAN as a practical and powerful solution for real-time TTA.
\end{onecolabstract}
]

\section{Introduction}

In today's competitive media environment, producing high-quality audiovisual content is essential. However, creating professional-grade sound typically requires expensive field or studio recordings, which drive up production costs and restrict creative endeavors for producers. Text-to-audio (TTA) generation can address these challenges by synthesizing sound directly from text, reducing production costs and streamlining workflows. Nevertheless, current TTA models face significant technical limitations including slow inference speeds and large model sizes that restrict their practical deployment and real-time applications, creating barriers to widespread adoption in professional media production pipelines.\\
Recent research in TTA has seen a surge of interest, with various approaches emerging to synthesize realistic and expressive audio from textual descriptions. Diffusion-based models \cite{diffsound, audioldm, make_an_audio, tango, make_an_audio2, audioldm2, tango2} achieve impressive quality by iteratively refining noisy latent representations, while autoregressive models like AudioGen \cite{audiogen} generate audio token by token to closely match the input text. However, diffusion-based methods require iterative denoising steps, resulting in high computational costs and slow inference speeds. Meanwhile, autoregressive approaches often suffer from error accumulation and large model sizes. These limitations underscore the need for more efficient and compact TTA models that can generate high-quality audio with rapid inference and low computational cost for practical deployment. \\
According to these requirements, we propose AudioGAN, a novel text-to-audio generation framework based on Generative Adversarial Networks (GANs) \cite{gan}. Unlike diffusion-based approaches that rely on iterative processing, AudioGAN generates audio in a single pass with a compact architecture. To address the inherent training challenges of GANs, we incorporated multiple contrastive losses into the training process. Moreover, to effectively embed textual information into the audio, we introduced innovative methods—namely, Single-Double-Triple (SDT) Attention and Time-Frequency Cross-Attention (TF-CA). We conducted various experiments to evaluate our method on the AudioCaps dataset. Our results show that AudioGAN not only achieves state-of-the-art performance but also significantly reduces model parameters and GPU memory usage. Notably, AudioGAN generates high-quality audio from text description in less than one second. To our knowledge, this is the first successful application of a GANs model in TTA, and the experimental results demonstrate AudioGAN a practical solution for real-time media production.

\section{Related Works}
Recent advances in TTA generation have led to a variety of approaches that aim to synthesize realistic and expressive audio directly from textual descriptions. Key milestones in this field include diffusion-based methods such as DiffSound \cite{diffsound}, AudioLDM \cite{audioldm, audioldm2}, Make-an-Audio \cite{make_an_audio, make_an_audio2}, and Tango \cite{tango, tango2}, which typically combine text-conditioned latent diffusion with VAE decoders \cite{stable_diffusion, vae}. These two-stage architectures have pushed the boundaries of audio fidelity by iteratively refining noisy latent representations. Additionally, autoregressive models like AudioGen \cite{audiogen} leverage sequential token prediction to generate audio that closely aligns with the input text. However, despite their impressive performance in terms of quality, diffusion-based approaches are hindered by their slow inference speeds—often requiring hundreds of iterative denoising steps even when techniques like DDIM \cite{ddim} are employed—while autoregressive models tend to suffer from error accumulation and are computationally intensive due to large model sizes \cite{autoregressive_error, transformer}.\\
\begin{figure}
\centering
\includegraphics[width=0.5\textwidth]{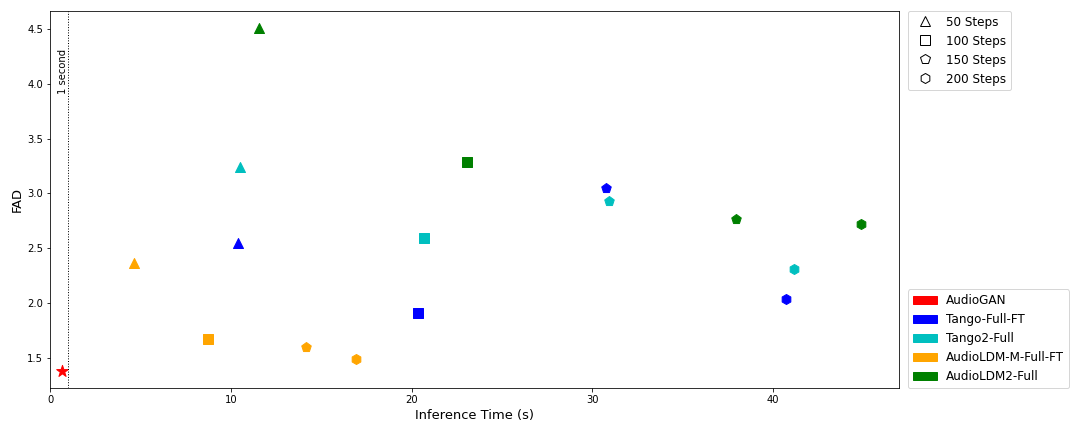}
\caption{Comparison of FAD scores (lower is better) and average inference times for various models on the AudioCaps test set. AudioGAN is represented by a red star, while diffusion-based models are depicted using different shapes to indicate their respective denoising steps, with colors differentiating individual models.}
\label{figure1}
\end{figure}
To address these challenges, our work departs from the conventional diffusion and autoregressive paradigms by employing a GANs \cite{gan} as the core generative model for TTA. By using a single generator model, our method eliminates the need for the multi-stage process that introduces performance gaps between separate components. This single-stage approach not only streamlines the generation process, significantly reducing both model size and computational overhead, but also enables rapid inference, making real-time application feasible.

\section{METHODS}
AudioGAN comprises three integral components: a text encoder utilizing the pre-trained Contrastive Language-Audio Pre-training (CLAP) model \cite{clap}, a generator constructed from GANs \cite{gan} trained via proposed methodology, and a vocoder employing the pre-trained HiFi-GAN \cite{hifi_gan}. The subsequent sections delineate the architectural design and training process of the GANs.

\begin{figure*}[h]
\centering
\includegraphics[width=0.9\linewidth]{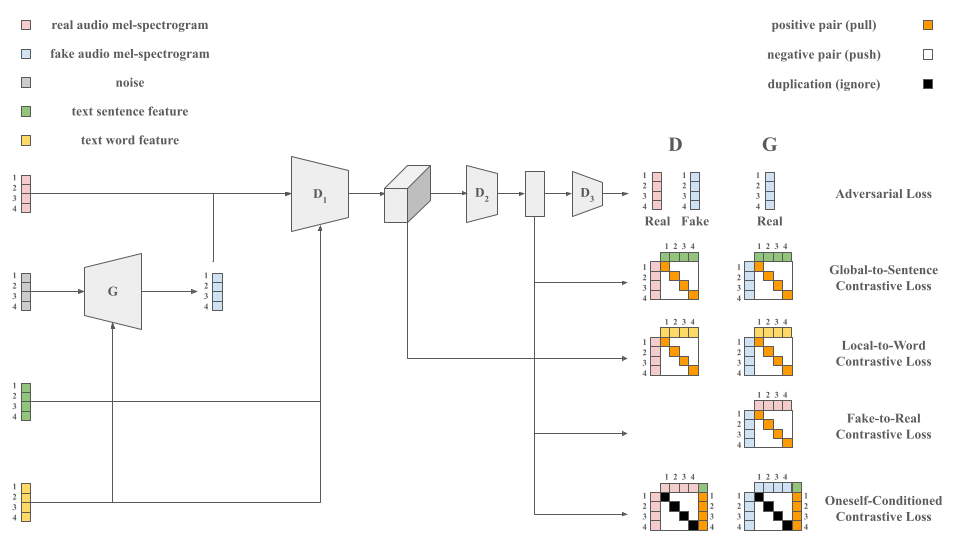}
\caption{A schematic of AudioGAN's training losses. The discriminator (D) optimizes four losses, while the generator (G) optimizes five losses. The audio local feature is first extracted from $ D_{1} $, then processed through $ D_{2} $ to obtain the global audio feature. Finally, the logit for the adversarial loss is extracted via $ D_{3} $.}
\label{figure2}
\end{figure*}

\subsection{The training of the GAN}
This section outlines our GANs training methodology. The generator synthesizes audio from textual features, while the discriminator leverages the text information to distinguish between real and synthesized audio. Multiple loss functions are employed to stabilize training and enhance performance, as illustrated in Figure 2.

\subsubsection{Adversarial Loss}
Within a mini-batch of $N$ audio-text pairs, let $i \in I \equiv \left\{ 1...N \right\}$ denote the index of a sample. Each audio sample is converted into a mel-spectrogram $m_{i}$, and its corresponding text is processed by a text encoder to extract word features $w_{i}$ and sentence features $s_{i}$. The generator $G$ then uses a noise vector $z_{i}$ together with these text features to produce a text-conditioned mel-spectrogram, while the discriminator $D$ evaluates both the generated and real mel-spectrograms for authenticity. We employ the hinge loss function \cite{hinge_loss} as the adversarial loss, with the objective functions for $D$ and $G$ detailed in the equations below.

\begin{footnotesize}
\begin{align} \label{eq1}
    \nonumber \mathcal{L}_{D} = & - \frac{1}{N} \sum_{i \in I} \min \left ( 0, -1 + D \left ( m_{i}, w_{i}, s_{i} \right )  \right) \\
    \nonumber                   & - \frac{1}{N} \sum_{i \in I} \min \left ( 0, -1 - D \left ( G\left( z_{i}, w_{i}, s_{i} \right ), w_{i}, s_{i} \right ) \right), \\
              \mathcal{L}_{G} = & - \frac{1}{N} \sum_{i \in I} D \left ( G\left( z_{i}, w_{i}, s_{i} \right ), w_{i}, s_{i} \right )
\end{align}
\end{footnotesize}

\subsubsection{Contrastive Loss}
Generated audio must accurately reflect the input text. To ensure this, we maximize mutual information by aligning global audio features with sentence-level text, local audio features with word-level text, and generated audio features with those from real audio sharing the same text \cite{xmc_gan}. We further enhance diversity by applying a oneself-conditioned contrastive loss that reduces similarity between distinct audio samples \cite{oc_supcon_gan}.

\noindent\textbf{Global-to-Sentence (G2S) Contrastive Loss.}
To enhance mutual information at the global level, we apply a contrastive loss between the audio global features $g$ and the sentence features $s$. These global features, extracted from the discriminator's penultimate layer for both real and generated audio, are encouraged to be similar for matching audio-text pairs while being pushed apart for non-matching pairs. In the equation below, $\cdot$ denotes the dot product, $\tau$ is the temperature parameter.

\begin{footnotesize}
\begin{align}\label{eq2}
     \mathcal{L}_{G2S} = - \frac{1}{N} \sum_{i \in I} \log \frac{\exp ( g_{i} \cdot s_{i} / \tau )}{\sum\limits_{ k \in I } \exp ( g_{i} \cdot s_{k} / \tau )}
\end{align}
\end{footnotesize}

\noindent\textbf{Local-to-Word (L2W) Contrastive Loss.}
Each word in the text includes significant meaning, and segments of the audio should reflect these words. To capture this without explicit annotations, we use an attention mechanism that aligns audio local features $l$ (extracted from the discriminator’s final convolutional layer) with word features $w$. First, soft attention $\alpha_{i,j}$ is computed between $l$ and $w$. Then, a contrastive loss based on a scoring function $\mathcal{S}(l, w)$ strengthens this alignment.

\begin{footnotesize}
\begin{align}\label{eq3}
     \alpha_{i,j} = \frac{ \exp ( \gamma_{1} ( w_{i} \cdot l_{j} ) )}{\sum_{k=1}^{R} \exp ( \gamma_{1} ( w_{i} \cdot l_{k} ) ) }, \;\, c_{i} = \sum_{j=1}^{R} \alpha_{i,j} l_{j}
\end{align}
\end{footnotesize}
\begin{footnotesize}
\begin{align}\label{eq4}
     \mathcal{S}(l, w) = \log \left ( \sum_{t=1}^{T} \exp ( \gamma_{2} ( w_{t} \cdot c_{t} ) ) \right ) ^ { \frac {1} {\gamma_{2}} }
\end{align}
\end{footnotesize}
\begin{footnotesize}
\begin{align}\label{eq5}
     \mathcal{L}_{L2W} = - \frac{1}{N} \sum_{i \in I} \log \frac{\exp ( \gamma_{3}  \mathcal{S}(l_{i}, w_{i}) )}{\sum\limits_{k \in I} \exp ( \gamma_{3}  \mathcal{S}(l_{i}, w_{k}) )}
\end{align}
\end{footnotesize}

Where $ R $ is the number of local features, $ T $ is the number of words, and $ \gamma_{1}, \gamma_{2}, \gamma_{3} $ are smoothing parameters.

\noindent\textbf{Fake-to-Real (F2R) Contrastive Loss.}
To produce high-quality audio, the generator can reference real audio samples. However, directly consulting real audio limit diversity. Instead, we encourage the generated audio to resemble its real counterpart in feature space by applying a contrastive loss between the global audio features of generated samples $g^{f}$ and those of real samples $g^{r}$. These features, extracted from the penultimate layer of the discriminator, guide the generator's learning without affecting the discriminator.

\begin{footnotesize}
\begin{align}\label{eq6}
    \mathcal{L}_{F2R} = -\log \frac{\exp ( g^{f}_{i} \cdot g^{r}_{i} / \tau )}{\sum\limits_{ k \in I } \exp ( g^{f}_{i} \cdot g^{r}_{k} / \tau )}
\end{align}
\end{footnotesize}

\begin{figure}
\centering
\includegraphics[width=0.45\textwidth]{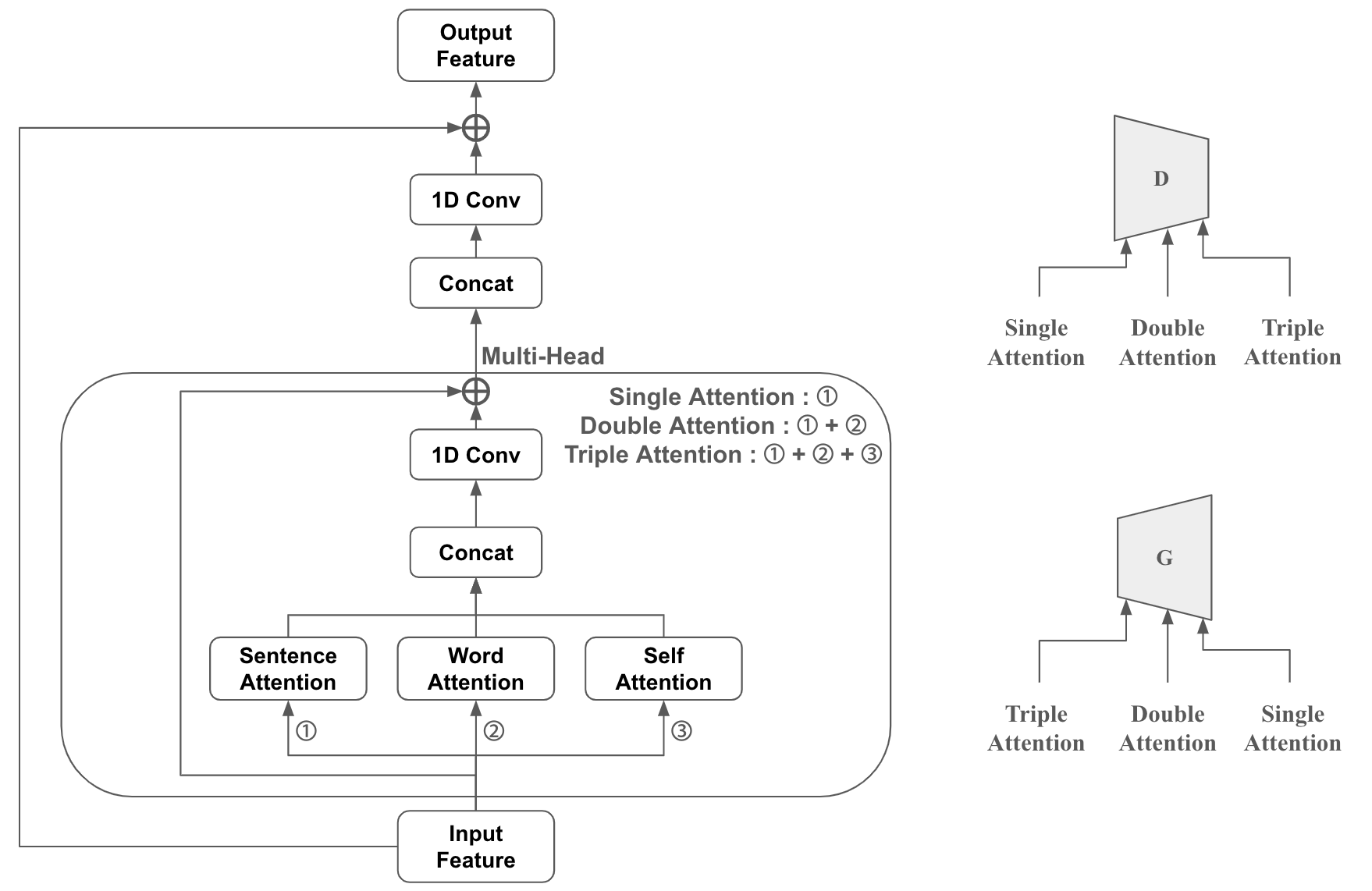}
  \caption{(Left) Flowchart of Single-Double-Triple (SDT) Attention and combination of attention mechanism at each stage. (Right) Schematic illustrating the application of SDT Attention within the discriminator and generator.}
\label{figure3}
\end{figure}

\noindent\textbf{Oneself-Conditioned Contrastive (OCC) Loss.}
The previous three contrastive losses ensure that each sample aligns with its corresponding text condition and that generated samples reference real data for enhanced quality. However, they focus solely on individual conditional alignment without regulating the relationships between different samples. As a result, even if text descriptions are similar, subtle differences may be lost, leading to redundant features and reduced diversity. To address this, we introduce an OCC Loss \cite{oc_supcon_gan} that pushes apart all samples while pulling each one closer to its respective sentence feature. This loss is applied to both generated and real samples, ensuring consistent regulation across the data distribution.

\begin{footnotesize}
\begin{align}\label{eq7}
    \mathcal{L}_{OCC} = - \frac{1}{N} \sum_{i \in I} \log \frac{\exp \left ( g_{i} \cdot s_{i} / \tau \right )}{\exp \left ( g_{i} \cdot s_{i} / \tau \right ) + \sum\limits_{ k \in I } {1}_{k \neq i}\exp \left ( g_{i}  \cdot g_{k} / \tau \right )}
\end{align}
\end{footnotesize}

\subsection{The architecture of the GANs}
This section describes the key components of our GANs model. To enhance text-conditioned audio generation, we integrate both self- and cross-attention mechanisms into the generator and discriminator. To balance performance with model complexity, we introduce a Single-Double-Triple Attention and a novel Time-Frequency Cross-Attention mechanism that effectively captures the temporal and spectral characteristics of audio data.

\subsubsection{Single-Double-Triple (SDT) Attention}
In our GANs framework, we employ both cross- and self-attention \cite{attention, transformer} in a complementary manner. Cross-attention is split into sentence attention, which captures the overall meaning of the text, and word attention, which focuses on fine-grained details. Self-attention captures long-range dependencies and intricate relationships within the audio feature. We introduce a hierarchical Single-Double-Triple (SDT) Attention scheme tailored to the distinct needs of the discriminator and generator, and each combination of attention is shown in Figure 3. In the discriminator, single attention captures global semantic content, double attention refines detailed alignment, and triple attention comprehensively assesses semantic authenticity. In the generator, triple attention is first performed to build a robust initial audio representation from noise and text, then gradually reduce complexity with double and single attention layers to refine the text-audio aligning and generate high-quality audio. This structured approach enables the discriminator to evaluate both global and fine-grained details and guides the generator to develop progressively refined audio representations.

\subsubsection{Time-Frequency Cross-Attention}

Conventional attention mechanisms \cite{image_attention, audio_attention} compute the query, key, and value from the same input feature, allowing every element to attend to every other element. However, in audio data, the two axes—time and frequency—encode distinct types of information. To leverage this structure, we separate audio features into two axes, enabling us to specialize in the unique characteristics of each. In our framework, we design novel attention mechanisms for self-attention and multi-modal cross-attention.

\begin{figure}
\centering
\includegraphics[width=0.45\textwidth]{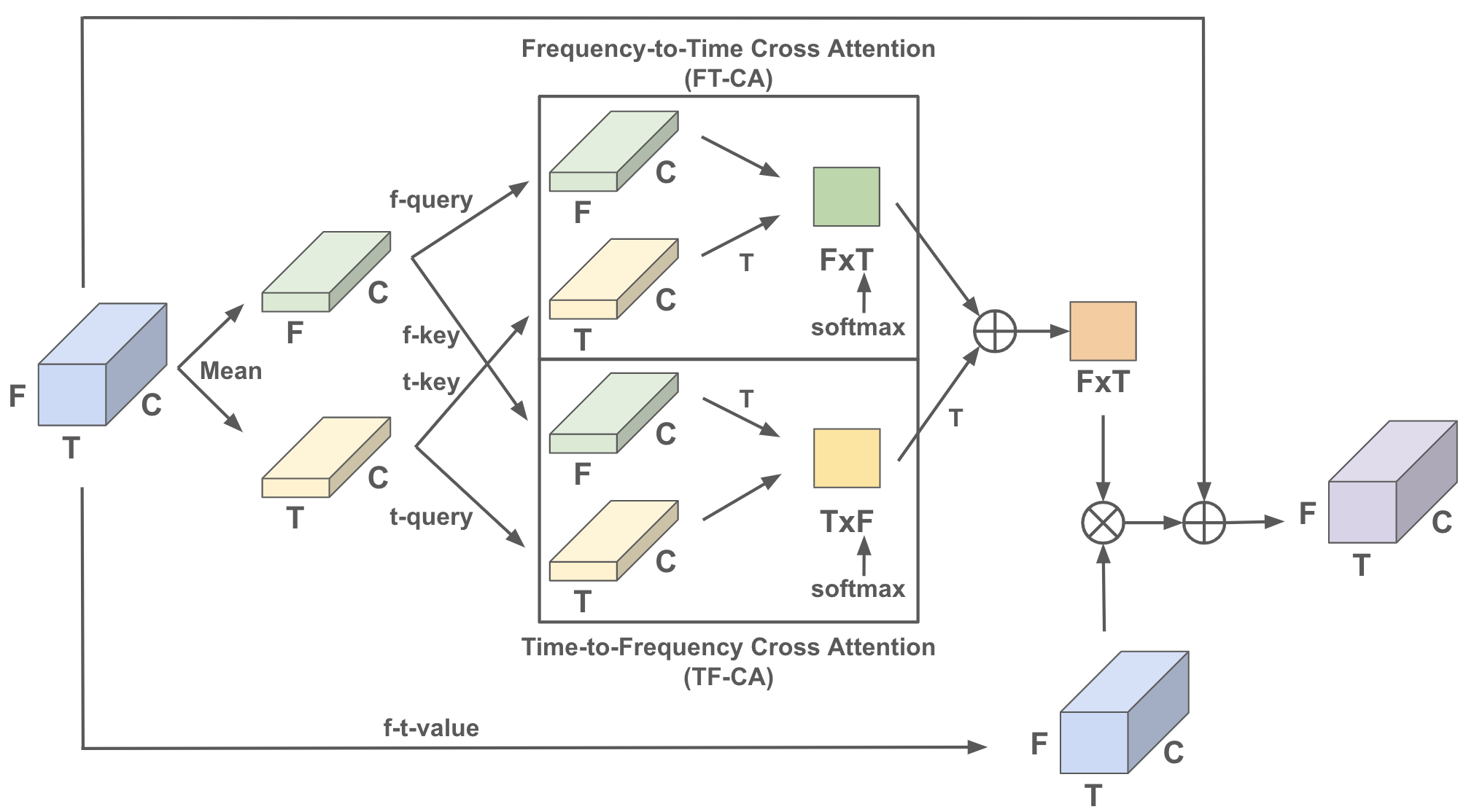}
  \caption{Illustration of the Self Time-Frequency Cross-Attention mechanism.}
\label{figure4}
\end{figure}

\textbf{Self Time-Frequency Cross-Attention}
Previous studies \cite{time_freq_attention_1, time_freq_attention_2, time_freq_attention_3} have applied attention separately to the time and frequency axes to model temporal and spectral cues, respectively. However, these methods do not account for the interdependency between these dimensions. We propose a self time-frequency cross-attention mechanism that decomposes the attention process into two complementary operations: frequency-to-time and time-to-frequency (Fig. 4).\\
The audio features from previous hidden layers are denoted by $x \in \mathbb{R}^{ F \times T \times C_{x} }$ where $F$ is the number of frequency bins, $T$ is the number of time steps, and $ C_{x} $ is the channel dimension. To derive frequency-specific and time-specific representations, we first average over the time and frequency dimensions, respectively:

\begin{figure}
\centering
\includegraphics[width=0.4225\textwidth]{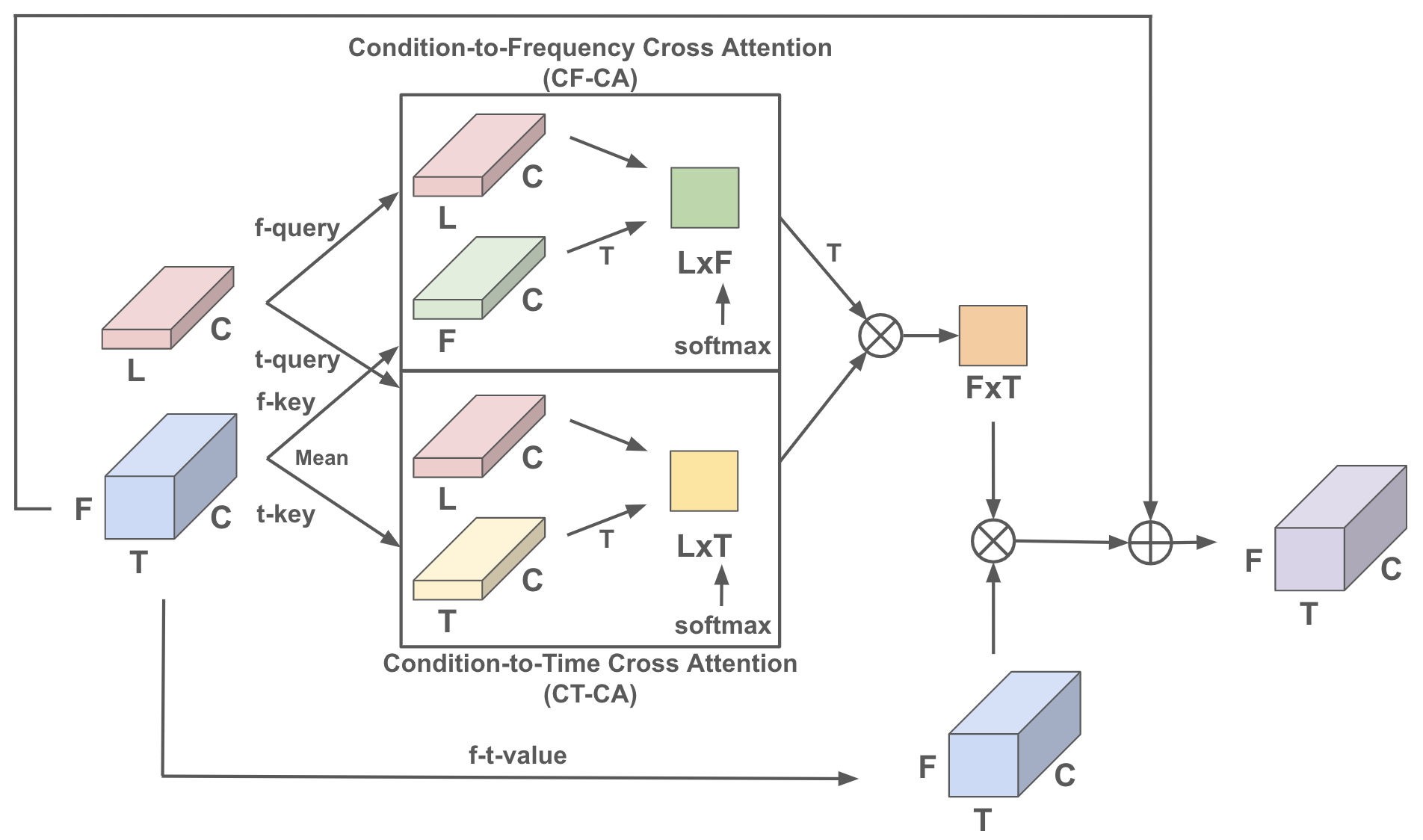}
  \caption{Illustration of  Multi Time-Frequency Cross-Attention mechanism.}
\label{figure5}
\end{figure}

\begin{align}\label{eq8}
\mathit{x}^f = \frac{1}{T} \sum_{t=1}^{T} \mathit{x} \left ( :, t, : \right ), \;
\mathit{x}^t = \frac{1}{F} \sum_{f=1}^{F} \mathit{x} \left ( f, :, : \right )
\end{align}

These aggregated features are projected into query and key spaces using trainable parameters $ \mathit{q}^{f}, \mathit{k}^{f}, \mathit{q}^{t}, \mathit{k}^{t} \in \mathbb{R}^{ C_{x} \times C_{h} } $ where $ C_{h} $ is the hidden channel dimension. In  frequency-to-time cross-attention, attention weights are computed from the frequency domain to capture temporal dynamics; each element of the resulting weight matrix quantifies the degree to which the $i$-th frequency feature attends to the $j$-th time feature:

\begin{align}\label{eq9}
\mathit{W}_{f2t} \left ( i, j \right ) = \frac{ \exp \left( \left ( x^{f}_{i} \cdot \mathit{q}^{f}_{i} \right ) \cdot \left ( x^{t}_{j} \cdot \mathit{k}^{t}_{j} \right )^{\top} \right ) } { \sum_{j=1}^{T} \exp \left( \left ( x^{f}_{i} \cdot \mathit{q}^{f}_{i} \right ) \cdot \left ( x^{t}_{j} \cdot \mathit{k}^{t}_{j} \right )^{\top} \right ) }
\end{align}

In contrast, time-to-frequency cross-attention computes attention weights in the time domain to capture spectral details. Each element represents the extent to which the $i$-th time feature attends to the $j$-th frequency feature:

\begin{align}\label{eq10}
\mathit{W}_{t2f} \left ( i, j \right ) = \frac{ \exp \left( \left ( x^{t}_{i} \cdot \mathit{q}^{t}_{i} \right ) \cdot \left ( x^{f}_{j} \cdot \mathit{k}^{f}_{j} \right )^{\top} \right )  } { \sum_{j=1}^{F} \exp \left( \left ( x^{t}_{i} \cdot \mathit{q}^{t}_{i} \right ) \cdot \left ( x^{f}_{j} \cdot \mathit{k}^{f}_{j} \right )^{\top} \right ) }
\end{align}

These attention maps are fused into a unified weights: 
\begin{align} \label{eq11}
\mathit{W} = \mathit{W}_{f2t} + \mathit{W}_{t2f}^{\top}
\end{align}
Finally, the output audio feature $ x' $ is obtained by element-wise multiplying the value feature—derived from the audio feature using a trainable parameter $ \mathit{v} \in \mathbb{R}^{ C_{x} \times C_{x} }$—with the combined attention weight  $ W $, followed by a residual connection:
\begin{align} \label{eq12}
\mathit{x}' = \mathit{x} + ( x \cdot v) \times \mathit{W}
\end{align}
By separating time and frequency, this mechanism effectively captures both temporal dynamics and spectral details, improving audio synthesis quality.

\begin{table*}[]
\caption{Comparison of AudioGAN with previous models on AudioCaps dataset.}
\fontsize{3}{4}\selectfont
\centering
\resizebox{\textwidth}{!}{%
\begin{tabular}{c|c|c|ccccc}
\hline
Model              & \makecell{Inference\\GPU Memory} & \#Param & FAD({$\downarrow$})  & FD({$\downarrow$})    & IS({$\uparrow$})   & KL({$\downarrow$})   & CLAP({$\uparrow$}) \\ \hline
AudioLDM-M-Full    & 10.25GB & 416M    & 3.84 & 32.2  & 5.82 & 1.67 & 0.45 \\
AudioLDM-M-Full-FT & 10.25GB & 416M    & 1.52 & 20.03 & 9.0  & 1.14 & 0.54 \\
AudioLDM-L-Full    & 15.87GB & 739M    & 3.69 & 34.69 & 7.77  & 1.56 & 0.42 \\
Tango              & 5.65GB  & 866M    & 1.44 & 15.09 & 7.41 & 1.08 & 0.54 \\
Tango-Full         & 5.65GB  & 866M    & 1.87 & 20.94 & 5.14 & 1.23 & 0.5  \\
Tango-Full-FT      & 5.65GB  & 866M    & 2.0  & 12.53 & 7.75 & 0.95 & 0.56 \\
AudioLDM2-Full     & 8.17GB  & 346M    & 2.74 & 18.5  & 8.08 & 1.25 & 0.46 \\
AudioLDM2-L-Full   & 10.95GB & 712M    & 2.92 & 20.29 & 7.25 & 1.22 & 0.46 \\
Tango2             & 5.65GB  & 866M    & 2.83 & 14.51 & 8.79 & 0.97 & 0.58 \\
Tango2-Full        & 5.65GB  & 866M    & 2.28 & 13.79 & 8.72 & 0.96 & 0.58 \\ \hline
AudioGAN           & 1.34GB  & 28M     & 1.38 & 12.66 & 9.63 & 1.15 & 0.49 \\ \hline
\end{tabular}%
}
\label{tab:1}
\normalsize
\end{table*}

\textbf{Multi Time-Frequency Cross-Attention} 
Beyond self attention, we extend our approach to condition the audio features on external textual information. Given an audio feature $x \in \mathbb{R}^{ F \times T \times C_{x} }$ and a condition feature $ c \in \mathbb{R}^{ N \times C_{c} }$ (where $ N $ represents the number of conditions—1 for sentence-level attention and the number of tokens for word-level attention), we first aggregate the audio features along the time and frequency dimensions as in Equation \eqref{eq8}. These aggregated representations, $ x_{f} $ and $ x_{t} $, along with the condition $ c $, are projected into query and key spaces using trainable parameters $ \mathit{q}^{f}, \mathit{q}^{t} \in \mathbb{R}^{ C_{c} \times C_{h} }$ and $ \mathit{k}^{f}, \mathit{k}^{t} \in \mathbb{R}^{ C_{x} \times C_{h} } $.\\
Unlike conventional cross-attention, where the main input is used as the query and the condition as the key, our method reverses this order: the condition serves as the query while each domain of the main input is treated as a key. This approach allows the model to evaluate the importance of time and frequency features from the condition's perspective, effectively emphasizing the most critical aspects of each domain. In the condition-to-frequency cross-attention, each element quantifies how strongly the $ i $-th condition feature attends to the $ j $-th frequency feature:

\begin{align} \label{eq13}
\mathit{W}_{c2f} \left ( i, j \right ) = \frac{ \exp \left( \left ( c_{i} \cdot \mathit{q}^{f}_{i} \right ) \cdot \left ( x^{f}_{j} \cdot \mathit{k}^{f}_{j} \right )^{\top} \right ) } { \sum_{j=1}^{F} \exp \left( \left ( c_{i} \cdot \mathit{q}^{f}_{i} \right ) \cdot \left ( x^{f}_{j} \cdot \mathit{k}^{f}_{j} \right )^{\top} \right ) }
\end{align}

Similarly, in the condition-to-time cross-attention, each element reflects the degree to which the $ i $-th condition feature aligns with the $ j $-th time feature:

\begin{align} \label{eq14}
\mathit{W}_{c2t} \left ( i, j \right ) = \frac{ \exp \left( \left ( c_{i} \cdot \mathit{q}^{t}_{i} \right ) \cdot \left ( x^{t}_{j} \cdot \mathit{k}^{t}_{j} \right )^{\top} \right ) } { \sum_{j=1}^{T} \exp \left( \left ( c_{i} \cdot \mathit{q}^{t}_{i} \right ) \cdot \left ( x^{t}_{j} \cdot \mathit{k}^{t}_{j} \right )^{\top} \right ) }
\end{align}
We then fuse these two attention maps using a dot product operation:

\begin{align} \label{eq15}
\mathit{W} = \mathit{W}_{c2f}^{\top} \cdot \mathit{W}_{c2t}
\end{align}
Finally, the updated audio feature $ x' $ is computed by applying the fused attention weights $ W $ to the value feature, followed by a residual connection (Eq. \eqref{eq12}).\\
This mechanism effectively aligns temporal and spectral audio features with textual conditions, ensuring fine-grained text details are accurately reflected in the generated audio.

\begin{figure*}[h]
\centering
\includegraphics[width=\linewidth]{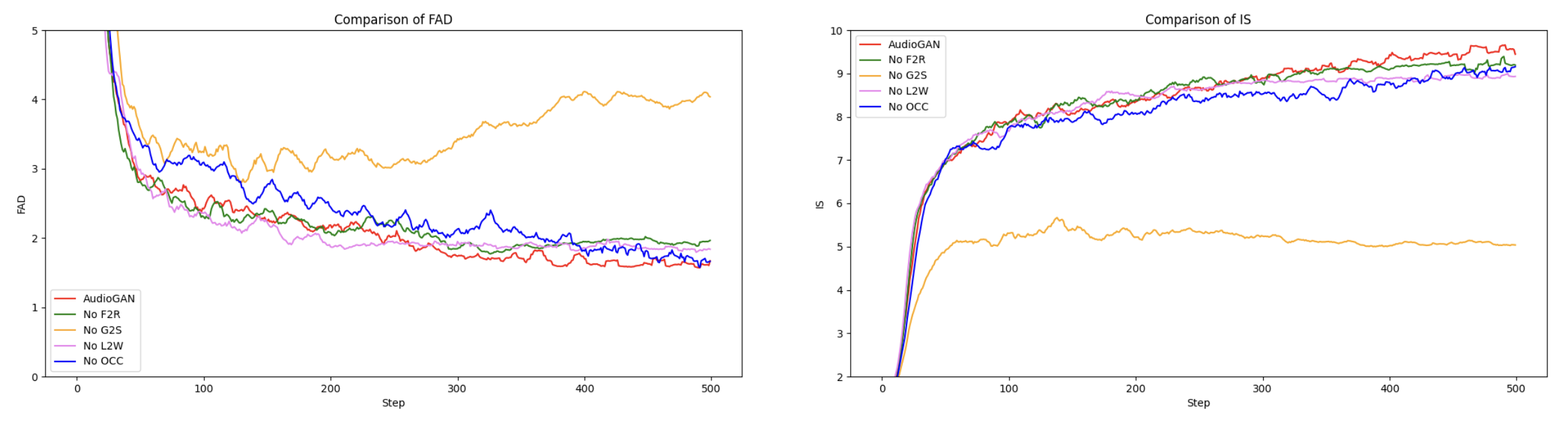}
\caption{Impact of removing the specific contrastive loss.}
\label{figure6}
\end{figure*}

\section{Experiment Setup}

\noindent\textbf{Dataset.} We used only the AudioCaps dataset \cite{audiocaps} for training and testing. We used approximately 49,000 clips for training and 964 clips for testing. Each sample is 10 seconds long with a 16 kHz sampling rate and was converted to mel-spectrograms (64 mel channels, 1024-sample window, 160-sample hop length) for training.

\noindent\textbf{Training Details.} The generator and discriminator learning rates were set to 0.0001 and 0.0004, respectively \cite{ttur}. The discriminator was updated twice per generator update. We use Adam optimizer \cite{adam} with $\beta_{1} = 0.5$ and $\beta_{2} = 0.999$ for training. For all contrastive losses, $\tau$ was set to 0.1, and for word contrastive loss, $\gamma_{1}$, $\gamma_{2}$, and $\gamma_{3}$ were set to 5, 5, and 50 \cite{attn_gan}, respectively.

\noindent\textbf{Evaluation Metrics.} We evaluated our generated audio using five key metrics: Frechet Audio Distance (FAD) \cite{fad}, Frechet Distance (FD) \cite{fd}, Inception Score (IS) \cite{is}, Kullback-Leibler Divergence (KL) \cite{kl}, and CLAP Score \cite{clap}. FAD compares the distribution of features from generated and real audio using the VGGish \cite{vggish} model, while FD does the same using the PANN model \cite{pann}. IS measures the quality and diversity of generated samples. KL divergence quantifies the difference between the distributions of real and generated audio based on a pre-trained classifier's output. Finally, the CLAP score assesses how well the generated audio aligns with the textual prompt. These metrics together provide a comprehensive evaluation of our model.

\section{Results}

\subsection{Comparison with Existing Methods}
Recent advances in TTA generation have been significantly advanced by diffusion-based models, such as AudioLDM \cite{audioldm, audioldm2} and Tango \cite{tango, tango2}, which rely on large U-Net architectures \cite{unet} to iteratively denoise latent representations. While these methods can achieve impressive audio quality with a sufficient number of denoising steps, they face a fundamental trade-off: increasing the number of steps improves fidelity but significantly escalates computational cost and inference time, whereas reducing the steps lowers computational cost and inference time but compromises audio quality. This limitation poses a major challenge for real-time applications such as interactive audio generation and on-the-fly sound design, where efficiency and responsiveness are critical.\\
In contrast, our method, AudioGAN, employs a GANs-based strategy for TTA. Instead of relying on an iterative inference process, AudioGAN generates audio in a single pass, significantly improving efficiency and reducing computational costs. In Table 1, we compare our model with existing methods on the AudioCaps test dataset, keeping the experimental conditions. Because generative models can produce slightly varied outputs from the same prompt, we averaged the results over three runs per approach. The results show that AudioGAN achieves state-of-the-art performance in key metrics, notably FAD and IS, and remains competitive in FD. The notably higher IS score indicates not only superior audio quality but also broader output diversity. Additionally, we measured the number of model parameters and GPU memory usage during inference. AudioGAN stands out with its compact architecture, significantly reducing both model size and memory requirements. As illustrated in Figure 1, while diffusion-based models improve performance with more diffusion steps at the cost of longer inference times, AudioGAN completes inference in under one second. These results highlight the advantages of our GANs-based strategy, which achieves competitive performance while remaining practical for real-time and resource-constrained TTA applications.

\begin{table}[]
\caption{Ablation results with different contrastive losses.}
\centering
\resizebox{7.5cm}{!}{
\begin{tabular}{c|cccc|ccc}
\hline
Method    & G2S          & L2W          & F2R          & OCC          & FAD($\downarrow$)  &  FD($\downarrow$)   &  IS($\uparrow$)  \\ \hline
No G2S &              & \checkmark   & \checkmark   & \checkmark   & 2.87 & 23.65 & 5.7  \\
No L2W & \checkmark   &              & \checkmark   & \checkmark   & 1.73 & 11.88 & 8.86 \\
No F2R & \checkmark   & \checkmark   &              & \checkmark   & 1.59 & 13.27 & 9.11 \\
No OCC & \checkmark   & \checkmark   & \checkmark   &              & 1.82 & 13.57 & 9.16 \\ \hline
AudioGAN & \checkmark   & \checkmark   & \checkmark   & \checkmark   & 1.38 & 12.66 & 9.63 \\ \hline
\end{tabular}
}
\label{tab:2}
\end{table}

\begin{table}[]
\caption{The ablation study of various attention.}
\centering
\resizebox{7.5cm}{!}{
\begin{tabular}{ccc|ccc}
\hline
\multicolumn{3}{c|}{Attention} & \multicolumn{1}{l}{} & \multicolumn{1}{l}{} & \multicolumn{1}{l}{} \\
First                & Second            & Third          & FAD($\downarrow$)  & FD($\downarrow$)    & IS($\uparrow$)   \\ \hline
self                 & self              & self           & 1.6  & 13.31 & 9.08 \\
word                 & word              & word           & 1.73 & 13.0  & 8.8  \\
sentence             & sentence          & sentence       & 2.0  & 12.53 & 9.01 \\ \hline
sentence & word, sentence    & self, word, sentence       & 1.38 & 12.66 & 9.63 \\ \hline
\end{tabular}
}
\label{tab:3}
\end{table}
\subsection{Ablation Study}
To verify the effectiveness of our proposed training techniques, we conducted comprehensive ablation studies targeting key components of our GANs training framework. First, we evaluated the contribution of each contrastive loss by removing one loss term at a time from the overall training objective and analyzing its impact on training stability and performance. Figure 6 and Table 2 show that excluding specific contrastive loss resulted in measurable changes in both stability and performance. Notably, the global-to-sentence (G2S) contrastive loss—which aligns the global audio features with the sentence-level text features extracted from the CLAP model—had the most pronounced impact. When the G2S loss was removed, the training became considerably unstable, and overall performance degraded significantly. This finding demonstrates that textual features aligned with auditory features are essential for effectively guiding model training in text-to-audio generation. Furthermore, our experiments show that incorporating the oneself-conditioned contrastive (OCC) loss led to a substantial improvement in the Inception Score (IS), indicating its effectiveness in promoting data diversity and enhancing the overall quality of generated audio. Collectively, these results highlight the critical role of each contrastive loss, ensuring stable training and improved synthesis performance. \\
We proposed a single-double-triple (SDT) attention that combines three different types of attention mechanisms (self, word, and sentence attention). To assess the individual impact of different attention mechanisms, we conducted experiments for the models trained using only one type of attention (Table 3). The results show that each individual attention mechanism provides decent performance, with self-attention yielding slightly higher results; however, self-attention incurs higher computational costs as feature sizes grow, compared to other methods. In contrast, SDT attention, which integrates all three mechanisms in a structured and balanced manner, enhances overall performance while mitigating the computational drawbacks of self-attention. We also acknowledge that numerous alternative combinations are possible, and exploring these variations remains a promising direction for future work.\\
Finally, to assess the effectiveness of our time-frequency cross-attention (TF-CA) mechanism, we conducted ablation studies comparing two variants: self TF-CA and multi TF-CA (Table 4). Self TF-CA was applied to the self-attention, and multi TF-CA was applied to the word and sentence attention, respectively. Our results show that incorporating self TF-CA yields modest performance gains over the baseline model without TF-CA. When multi TF-CA is used, both FAD and FD scores improve slightly, while the IS score increases modestly. Notably, combining both variants leads to even greater improvements. These results demonstrate that our TF-CA mechanism effectively captures the unique temporal and spectral characteristics of audio, thereby enhancing both the quality and diversity of the generated outputs.

\begin{table}[]
\caption{The ablation study of Time-Frequency Cross Attention (TF-CA).}
\centering
\resizebox{5.5cm}{!}{
\begin{tabular}{cc|ccc}
\hline
\multicolumn{2}{c|}{TF-CA} & \multirow{2}{*}{FAD($\downarrow$)} & \multirow{2}{*}{FD($\downarrow$)} & \multirow{2}{*}{IS($\uparrow$)} \\
Self       & Multi      & & & \\ \hline
           &            & 1.76 & 13.82 & 8.62 \\
\checkmark &            & 1.67 & 13.68 & 8.86 \\
           & \checkmark & 1.86 & 14.42 & 9.05 \\ \hline
\checkmark & \checkmark & 1.38 & 12.66 & 9.63 \\ \hline
\end{tabular}
}
\label{tab:4}
\end{table}

\section{Conclusion}
This paper propose the first successful attempt at using a GANs-based framework for TTA generation. Our approach, AudioGAN, overcomes the inherent challenges of diffusion-based methods—such as large model sizes, slow inference, and high computational costs—by generating audio in a single pass with fewer parameters in under one second. To achieve this, we incorporate multiple contrastive losses and introduce novel methods, namely Single-Double-Triple (SDT) Attention and Time-Frequency Cross-Attention (TF-CA). Extensive experiments on the AudioCaps dataset demonstrate that AudioGAN not only achieves state-of-the-art performance but also significantly reduces model parameters and GPU memory usage. These results highlight the advantages of our GANs-based approach in generating high-fidelity and diverse audio while maintaining computational efficiency, thereby paving the way for real-time applications and advancing the field of TTA.

\bibliographystyle{jaes}


\end{document}